\documentstyle[preprint,aps]{revtex}
\begin{document}
\preprint{}
\title{Initial Data for General Relativity\\
with Toroidal Conformal Symmetry
}
\author{R. Beig and S. Husa}
\address{Institut f\"ur Theoretische Physik,
         Universit\"at Wien, Vienna, Austria}
\date{\today}
\maketitle
\begin{abstract}
A new class of time-symmetric solutions to the initial value constraints of
vacuum General Relativity is introduced. These data are globally regular,
asymptotically flat (with possibly several asymptotic ends) and in general
have no isometries, but a $U(1)\times U(1)$ group of conformal isometries.
After decomposing the Lichnerowicz conformal factor in a double Fourier series
on the group orbits, the solutions are given in terms of a countable family of
uncoupled ODEs on the orbit space.
\end{abstract}
\pacs{PACS 4.20.Ex}
\narrowtext
%

If one casts General Relativity in canonical form one finds that the Einstein
equations split into two parts.
One part, the constraints, has no time derivatives in it: the equations are
conditions on the initial data. The other part, the evolution equations,
determine how this initial data develop in time.
The constraints are a highly underdetermined system which can be turned
into elliptic equations by a judicious choice of the free data.

The simplest version of the constraints occurs in the so-called
moment-of-time-symmetry (or time-symmetric) problem, where the velocity part
of the data, namely the extrinsic curvature of the initial 3-manifold $M'$, is
set to zero and the only remaining equation is $R[g]=0$, with $g$ a Riemannian
metric on the Cauchy slice and $R$ the scalar curvature of $g$.

The standard approach, pioneered by Lichnerowicz,
is to select a base metric $g$ and construct
a solution metric $g'$ by means of a conformal transformation $g'=\psi^{4}g$.
One then gets a linear elliptic PDE, namely
\begin{equation}\label{time-sym-eq}
L_{g}\psi:=(-\bigtriangleup_{g}+ \frac{1}{8}R[g])\psi = 0,\qquad \psi > 0
\end{equation}
for the conformal factor. In the asymptotically flat case studied here one
in addition assumes that $g'$ tends to the flat metric at infinity.
One then requires that $\psi$, in addition to obeying Eq. (\ref{time-sym-eq}),
go to one in that limit.

Simplification should occur if one imposes some symmetry on $g$.
Completely reducing this 3 dimensional PDE problem to a one dimensional ODE
by imposing a symmetry with 2 dimensional orbits does not really work.
In the case of spherical symmetry this would, by the Birkhoff theorem, leave
us with data for the Schwarzschild solution. In the cylindrically symmetric
case, using the boundary condition, only Minkowski space remains \cite{BCM}.
Thus, in the standard approach, the simplest solution that has
been identified is to assume that the free data is axially symmetric which
reduces Eq. (\ref{time-sym-eq}) to a two dimensional PDE, and much effort,
especially by the numerical community, has been devoted to solving it.

In this paper we wish to introduce a new symmetry in the base metric which
gives us a large class of non-trivial solutions while at the same time allowing
us to benefit from the underlying linearity of the equation. With this symmetry
we can reduce the equation to a set of uncoupled linear ODEs. This is a
remarkable simplification which offers major benefits in a whole range of
problems. The key idea is to use the freedom in choosing the base metric $g$
in a conformal equivalence class, particularly one can choose a base
metric that lives on a compact manifold which can be opened up to an
asymptotically flat one by conformal decompactification (analogous to
stereographic projection).

The physical data we are considering are data defined
on a 3-dimensional manifold $M'$ which may have several ends and are
asymptotically flat near each of them. They possess two
commuting, orthogonal conformal Killing vector fields. Their action extends to
a $U(1)\times U(1)$ action on the many-point compactification $S^3$ of $M'$.
After a suitable conformal rescaling of the metric, this action becomes in
fact an isometry, and is thus of the ``polarized Gowdy type'' \cite{Chrusciel}.
In the special case of just one point-at-infinity, which is a fixed point of
one $U(1)$ factor, these data have this $U(1)$ factor as an isometry and
are thus a subclass of Brill waves. The fact that the group of all conformal
symmetries of the physical data extends to an isometry group only after
conformal compactification turns out to be a blessing rather than a drawback.

Even in the absence of symmetry it is useful to replace Eq.
(\ref{time-sym-eq}), on $R^{3}$ say, together with the boundary condition
$\psi\rightarrow 1$ at infinity, by the equation
\begin{equation}\label{constraint}
L_{g}G=4\pi\delta_{\Lambda},\qquad G>0
\end{equation}
on the one-point compactification $S^{3}$ of $M'$ with $g$ now being a
Riemannian metric on $S^{3}$ and $\delta_{\Lambda}$ the
Dirac delta function concentrated at some point $\Lambda\in S^3$.
It is then easy to see, using the known asymptotic behaviour of $G$ near
$\Lambda$, that the metric $g'_{ab}=G^{4}g_{ab}$ on
$R^{3}\cong S^{3}\backslash\Lambda$ is asymptotically flat near $\Lambda$
and satisfies $R[g']=0$ on account of Eq. (\ref{constraint}).
A solution $G>0$ is known to exist, iff $\lambda_{1}$, the lowest eigenvalue of
$L_{g}$,
is positive \cite{beig+om}. In particular, this existence is insensitive to the
choice of the point $\Lambda$ which we will use later when we construct
solutions with many asymptotic ends.

To solve Eq. (\ref{constraint}) on the compact manifold $S^{3}$ we subtract
from $G$ a quantity $G_{loc}$, taken to
be a suitable approximation to the Hadamard fundamental solution
\cite{garabedian} of $L_{g}$
near $\Lambda$, arbitrarily extended to all of $S^{3}$. By including a
sufficient number of terms in $G_{loc}$, we can arrange for the r.h.s. in
\begin{equation}\label{smoothed}
L_{g}\phi=L_{g}(G-G_{loc})=\rho_{\Lambda}
\end{equation}
to be as smooth at $\Lambda$ as we please. Solving the regular equation
(\ref{smoothed}) on $S^{3}$ numerically the usual problems arising from
having to use a finite grid for an infinite region ($R^{3}$ in this case)
are avoided. Moreover, in the presence of a $U(1)\times U(1)$ isometry
we can introduce adapted coordinates, which will allow us to reduce
the PDE (\ref{smoothed}) to a system of uncoupled ODEs.

For the basic setup it is best to start by obtaining the standard metric
$\raisebox{-0.5mm}{$\stackrel{o}{g}$}_{ab}$ on $S^{3}$  by inverse
stereographic projection from $R^{3}$ with flat metric
$\raisebox{-0.5mm}{$\stackrel{o}{g}$}{}'_{ab}$. Let $z,r,\varphi$ be standard
cylindrical coordinates on $(R^{3},\raisebox{-0.5mm}{$\stackrel{o}{g}$}{}')$.
Taking $\Omega=4(1+4(r^{2}+z^{2}))^{-1}$, the metric
$\raisebox{-0.5mm}{$\stackrel{o}{g}_{ab}$}=
\Omega^{2}\raisebox{-0.5mm}{$\stackrel{o}{g}$}{}'_{\!ab}$
extends to the standard one on $S^{3}$. Furthermore
$\eta^{a}=(\frac{\partial}{\partial\varphi})^{a}$ extends
to a Killing vector on $(S^{3},\raisebox{-0.5mm}{$\stackrel{o}{g}$}_{\!ab})$.
With $\rho^{2}=\eta^{a}\eta^{b}\raisebox{-0.5mm}{$\stackrel{o}{g}$}_{ab}$
we have that $\eta^{a}\rho_{a}=0$ and
\begin{equation}\label{de}
\raisebox{0.1mm}{$\stackrel{o}{D}$}{}_{a}\raisebox{-0.5mm}{$\stackrel{o}
{\eta}$}_{b}=
2\rho^{-1}\rho_{[a}\raisebox{-0.5mm}{$\stackrel{o}{\eta}$}{}_{b]},
\qquad\rho_{ab}=
\rho^{-3}\raisebox{-0.5mm}{$\stackrel{o}{\eta}$}_{a}\raisebox{-0.5mm}
{$\stackrel{o}{\eta}$}_{b} - \rho \raisebox{-0.5mm}{$\stackrel{o}{g}$}_{ab},
\end{equation}
where we use the definitions
\begin{displaymath}
\raisebox{0.1mm}{$\stackrel{o}{D}$}{}_{c}
\raisebox{-0.5mm}{$\stackrel{o}{g}$}_{ab}=0,
\qquad
\rho_{a}=\raisebox{0.1mm}{$\stackrel{o}{D}$}{}_{a}\rho, \qquad
\rho_{ab}=\raisebox{0.1mm}{$\stackrel{o}{D}$}{}_{a}
\raisebox{0.1mm}{$\stackrel{o}{D}$}{}_{b}\rho.
\end{displaymath}
Next define
\begin{equation}\label{defxi}
\xi^{a}=\rho^{-1}\raisebox{-0.5mm}{$\stackrel{o}{\epsilon}$}{}^{abc}\rho_{b}
\raisebox{-0.5mm}{$\stackrel{o}{\eta}$}{}_{c},
\end{equation}
obeying
$\xi^{a}\xi^{b}\raisebox{-0.5mm}{$\stackrel{o}{g}$}_{ab}=1-\rho^2=:\sigma^2$,
$\xi^{a}\rho_{a}=0$,
$\xi^{a}\eta^{b}\raisebox{-0.5mm}{$\stackrel{o}{g}$}_{ab}=0$, and
\begin{equation}\label{dxi}
\stackrel{o}{D}_{a}\xi_{b}=
2\sigma^{-1}\sigma_{[a}\xi_{b]}.
\end{equation}

It thus follows from equations (\ref{de},\ref{defxi},\ref{dxi})
that $(\xi,\eta)$ form a pair of commuting, orthogonal
and hypersurface-orthogonal Killing vectors of
$\raisebox{-0.5mm}{$\stackrel{o}{g}$}_{ab}$ spanning
$\rho=\rho_{0}=const.$ where $\rho_{0}$ ranges between $0$ and $1$. For
$\rho_{0}\neq 0,1$ these are flat tori of constant mean curvature changing sign
at the Clifford torus for which $\rho_{0}=\frac{1}{\sqrt 2}$. The sets $\rho=0$
($\rho=1$) are closed lines on which $\eta^{a}$ ($\xi^{a}$) is zero (the
``axes''). In fact, $\rho=0$ (respectively $\rho=1$) are linked great circles
on $S^{3}$,
corresponding to the $z$-axis (respectively the circle $z=0,r=1/2$) after
stereographic projection. Using coordinates $(\rho,\chi,\varphi)$ where
$\xi^{a}=(\frac{\partial}{\partial\chi})^{a}$ and $\chi=const.$ are orthogonal
to $\xi^{a}$ we find that $\chi$, like $\varphi$, has periodicity $2\pi$.
The coordinate $\chi$ can be fixed by requiring $\Lambda$, the point at
infinity under stereographic projection, to be at $\rho=0$, $\chi=0$.
The totally geodesic surfaces $\chi=const.$ ($\varphi=const.$) are metric
hemispheres intersecting at $\rho=1$ ($\rho=0$) with
$\{\chi=\alpha\}\cup\{\chi=\alpha+\pi\}$
($\{\varphi=\beta\}\cup\{\varphi=\beta+\pi\}$) being
smoothly embedded $S^{2}$'s.
Under stereographic projection the surfaces $\chi=const.$ do not change
topology
except for $\{\chi=0\}\cup\{\chi=\pi\}$ which gets decompactified into
$\{z=0, r\geq 1/2\}\cup\{z=0, r\leq 1/2\}$, i.e. the equatorial plane $z=0$.

In the $(\rho,\chi,\varphi)$ coordinates the metric
$\raisebox{-0.5mm}{$\stackrel{o}{g}$}_{ab}$ takes the form
\begin{equation}\label{ds0}
\raisebox{-0.5mm}{$\stackrel{o}{g}$}_{ab}=\frac{\rho_{a}\rho_{b}}{1-\rho^{2}}
+ (1-\rho^{2})\chi_{a}\chi_{b} + \rho^{2}\varphi_{a}\varphi_{b}.
\end{equation}
We now generalize Eq. (\ref{ds0}) by setting
\begin{equation}\label{ansatz}
g_{ab}=e^{Aq}\left[\frac{\rho_{a}\rho_{b}}{1-\rho^{2}}
+ (1-\rho^{2})\chi_{a}\chi_{b}\right] +  \rho^{2}\varphi_{a}\varphi_{b}
\end{equation}
where $A$ is a constant and $q$ is a smooth function of $\rho^{2}$ with
$q\vert_{\rho=0}=0$. Eq. (\ref{ansatz}) then defines a smooth metric on
$S^{3}$ which, again, has a $U(1)\times U(1)$ symmetry with the same properties
as above, with the exception that the hemispheres $\chi=const.$ and
$\varphi=const.$ are not now metric half-spheres and that maximal tori
$\rho=\rho_{0}$ occur whenever
\begin{displaymath}
A\rho(1-\rho^{2})q'+1-2\rho^{2}=0,\qquad \mbox{where}\qquad
q'=\frac{dq}{d\rho}.
\end{displaymath}

The metrics defined by (\ref{ansatz}) are the free data we wish to consider.
The first step now consists of checking whether $\lambda_{1}(g) > 0$.
This is known to be true
for $A_{1}<A<A_{2}$ \cite{shusa}, where $A_{1}<{0}$, $A_{2}>0$ are numbers
dependig on $q$, both finite provided $q\not\equiv 0$.
In the present case $\lambda_{1}(A)$ is determined by an ODE: by virtue of
non-degeneracy of ground states, the corresponding eigenfunction has to share
the $U(1)\times U(1)$-symmetry  of the problem and is a function merely of
$\rho$. (In contrast, in Eq. (\ref{constraint}), the original symmetry is
broken by one's choice of $\Lambda$.)
It is known that for $A$ sufficiently close to $A_{1}$ or $A_{2}$, the
constant term in $G$, i.e. the ADM mass gets arbitrarily large and the physical
metric $g'$ develops trapped surfaces (see Ref. \cite{beig+om}).

In the next step we subtract a suitable $G_{loc}$ from $G$ and try to solve Eq.
(\ref{smoothed}). We can now expand $\phi$ and $\rho_{\lambda}$ in a
Fourier series on the square torus ($\chi\in[0,2\pi]$,$\varphi\in[0,2\pi]$).
Since $L_{g}$ commutes with $\frac{\partial}{\partial \chi}$ and
$\frac{\partial}{\partial \varphi}$ this results in an ODE on $\rho\in[0,1]$
for each Fourier mode, supplemented by boundary conditions at $\rho=0$
and $\rho=1$ needed to ensure regularity of $\phi$ on the two axes.
By uniqueness \cite{CKV} an isometry of $g_{ab}$
carries over to $g'_{ab}$ iff $\Lambda$ is a fixed point of this isometry.
Thus, if $\Lambda$ is in a general position, $g'$ has no isometries
whatsoever. On the other hand, taking $\Lambda$ to be at $\rho=0$, $\chi=0$,
say, we keep $\frac{\partial}{\partial\varphi}$ as a Killing vector, and the
resulting data are a special case of Brill waves \cite{brill}. The vector field
$\frac{\partial}{\partial\chi}$ survives on $R^{3}=S^{3}\backslash\Lambda$
merely as a conformal vector field which is incomplete along the axis $\rho=0$.
There is however the discrete ``mirror'' isometry $(\rho,\chi,\varphi)
\mapsto(\rho,2\pi-\chi,\varphi)$, leaving the equator invariant.

We can generalize the above to the case where there are several asymptotic
ends. We simply choose a finite number of points $\Lambda_{i}$ $(i=1,\dots,N)$,
solve numerically for the corresponding Green function $G_{i}$ and write
\begin{equation}\label{superposition}
g'_{ab}=\left(\sum_{i=1}^{N}c_{i}G_{i}\right)^{4}g_{ab},
\end{equation}
where the ``source strengths'' $c_{i}$ are arbitrary positive numbers. When
$A=0$, these are simply the many-black-hole solutions discussed  by Brill
and Lindquist \cite{B+L}.

If all $\Lambda_{i}$ lie on the two axes,
each $G_{i}$ in $G=\sum_{i=1}^{N}c_{i}G_{i}$ corresponds to an
axially symmetric (Brill-type) solution. It is perhaps
useful to observe that Brill's simple positive-mass proof \cite{brill}
immediately gives positivity of mass also in this non-axially symmetric,
topologically nontrivial situation.
Another useful observation, in the case of a general distribution
of $\Lambda_{i}$'s, is that not all of the $G_{i}$'s have to be
computed separately:
for any $\Lambda_{1}$, $\Lambda_{2}$ lying on the same orbit of the isometry
group, $G_{2}$ is given in terms of $G_{1}$ by a product of suitable
rotations in $\varphi$ and $\chi$.
Take e.g. $N=2$ with $\Lambda_{1}$ at $(\rho=0,\chi=0)$ and $\Lambda_{2}$ at
$(\rho=0,\chi=\alpha)$. In the special case $\alpha=\pi$, the physical metric
$g'_{ab}$ has the mirror symmetry $\chi\mapsto2\pi-\chi$. When $c_{1}=c_{2}$,
it has the inversion symmetry
$\{\chi\}\cup\{\chi+\pi\}\mapsto\{\alpha-\chi\}\cup\{\alpha-\chi+\pi\}$,
 leaving
the ``throat'' $\{\chi=\alpha/2\}\cup\{\alpha/2+\pi\}$ invariant. Thus the
``throat'' is totally geodesic with respect to $g'_{ab}$, in particular a
minimal embedded 2-sphere (a ``horizon''). We remark that, when $A=0$, the
above discrete symmetries are present irrespective of $c_{1}$, $c_{2}$ and
$\alpha$. Because, then, by combining a homothety with a proper conformal
motion coming from the Killing vector $\frac{\partial}{\partial z}$ on $(R^{3},
\raisebox{-0.5mm}{$\stackrel{o}{g}$}{}'_{ab})$,
we can always arrange for $c_{1}'=c_{2}'$, $\alpha'=\pi$. The corresponding
physical data is of course nothing but a time-symmetric slice of the Kruskal
spacetime with mass
$m(c_{1},c_{2},\alpha)=2\sqrt{2}c_{1}c_{2}(1-\cos{\alpha})^{-1/2}$.
When $A\not=0$ our solutions for $c_{1}\not=c_{2}$ are not
inversion symmetric. Placing sources of equal strength at $\rho=0$,
$\alpha_{i}=\frac{2\pi i}{N}$, $i=0,1,\dots,N-1$, we find in an analogous
manner that on $R^{3}\backslash\bigcup_{i=1}^{n}\Lambda_{i}$ each asymptotic
end is surrounded by a throat. Thus, viewing $\Lambda_{1}$ as ``infinity'' and
the
other $\Lambda_{i}$'s as ``black holes'' or ``particles'' we can in particular
say that these objects, viewed from infinity, are so close together that,
in addition to a horizon for each of them individually, there is one
surrounding them all.

Finally we want to place the solutions found here in a more general context.
 From the conformal-compactification viewpoint adopted here, looking for
time-symmetric initial data with symmetries naturally leads to looking for
metrics $g$ on a compact 3-manifold $M$, not necessarily $S^{3}$,
with a group $\cal G$ of conformal
symmetries. Suppose $(M,g)$ is not conformally diffeomorphic to $(S^{3},
\raisebox{-0.5mm}{$\stackrel{o}{g}$})$
(in which case the one-point conformal decompactification would be data for
Minkowski spacetime. Then, by virtue of a theorem due to Obata \cite{obata},
$\cal G$ acts by isometries on a metric $g_{1}$ in the conformal class of $g$.
Thus, calling $g_{1}$ again $g$, we are left with isometries of $(M,g)$.
When $\cal G$ has 3-dimensional orbits, it has to be one of the Bianchi types
existing on a compact manifold. Whether they too can be usefully employed as
background metrics for conformal decompactification, we have not investigated.
The next case is the one of 2-dimensional orbits which has been fully
classified \cite{Chrusciel2}. Taking $M$ to be $S^{3}$, which is the
topologically trivial case in the present context, $\cal G$ can be $SO(3)$
or $U(1)\times U(1)$. The first case is again conformally, whence physically,
trivial. The second case, provided the two Killing vectors can be chosen to
be orthogonal, is the polarized Gowdy case, which is conformally diffeomorphic
to the one used here.

If $M$ is chosen to be $S^{2}\times S^{1}$ and ${\cal G}=SO(3)$, $(M,g)$ is
conformally equivalent to $S^{2}\times S(a)$, the standard $S^{2}$ times the
circle of radius $a$. This, in physical terms, corresponds to the
Misner wormhole \cite{misner}.
The case ${\cal G}=U(1)\times U(1)$ is a generalization of the
Misner wormhole, and is similar to the case studied here.

Finally, when $M\cong T^{3}$, it follows from a fundamental theorem of Schoen
and Yau \cite{SY} that $\lambda_{1}(g)\leq 0$ for all metrics $g$, so that no
positive Green function exists.

Let us summarize: We have presented here a class of initial data for General
Relativity which in general have no symmetries whatsoever and for which the
constraints are equivalent to an infinite system of uncoupled ODEs.
In forthcoming work by one of us (S.H.), the procedure outlined here is carried
out for some classes of initial data having the required $U(1)\times U(1)$
conformal symmetry. This will involve a numerical study of the existence
question, i.e. the sign of $\lambda_{1}(g)$, and  the  actual construction of
initial data. Also included  will be the location of apparent horizons and
tests for the isoperimetric inequality of black holes and the hoop conjecture.
Since our method is fundamentally different from the one usually
employed in Numerical Relativity based on solving PDEs directly on 2 or 3
 dimensional
grids, we expect  our solutions to be useful testbeds for ones obtained by
these standard methods. Clearly our method so far makes decisive use of
the linearity of the Lichnerowicz equation (\ref{constraint}) in the
time-symmetric case. In the future we plan to study the nonlinear case of
maximal non-time-symmetric initial data having a $U(1)\times U(1)$ symmetry
of the background metric and background extrinsic curvature. Whether the lines
of thought followed in the present work offer simplifications also for the
evolution problem is at present unclear.

This work was supported by Fonds zur F\"orderung der
wissenschaftlichen Forschung, Project No. P09376-PHY.

\end{document}